\documentstyle[preprint,aps]{revtex}
\tightenlines

\begin{document}
\draft
\title{ Inertial Mass of a Vortex in Cuprate Superconductors}
\author{D.M. Gaitonde$^\dagger$\\}
\address{Jawaharlal Nehru Centre for 
          Advanced Scientific Research\\ 
          Bangalore 560 064\\
             INDIA \\ and \\ }

\author{T.V. Ramakrishnan $^{* \S}\/$\\}
\address{Department of Physics\\ Indian Institute of Science\\
  Bangalore 560 012\\
  INDIA\\}
\maketitle
\date{\today}
\begin{abstract}
We present here a calculation of the inertial mass of
a moving vortex in cuprate superconductors.
This is a poorly known basic quantity of obvious
interest in vortex dynamics.  The motion of a vortex
causes a dipolar density distortion and an associated electric field
which is screened. The energy cost of the density distortion as well as
the related screened electric field contribute to the
vortex mass, which is small because of efficient
screening.  As a preliminary, we present a discussion
and calculation of the vortex mass using a microscopically derivable
phase-only action functional for the far region
 which 
 shows that the contribution from the far region is negligible,
 and that most of it arises 
 from the (small) core region of the vortex.  
A calculation based on a phenomenological Ginzburg-Landau functional 
is performed  in the core region. 
Unfortunately such a  calculation is unreliable, 
the reasons for it are discussed.
A credible calculation of the vortex mass thus requires a
fully microscopic, non-coarse grained theory.  This is
developed, and results are presented for a s-wave BCS
like gap, with parameters appropriate to the cuprates.
The mass, about 0.5 $m_e\/$ per layer, for 
magnetic field along the $c\/$ axis, arises from
deformation of quasiparticle states bound in the core,
and screening effects mentioned above.  We discuss
earlier results, possible extensions to d-wave symmetry, and observability
of effects dependent on the inertial mass.
\end{abstract}
\pacs{PACS numbers: 74.20-z; 74.20.Fg; 74.20.De }

\noindent $^\dagger\/$ Present and permanent address:- Mehta
Research Institute, 10, Kasturba Gandhi Marg, Allahabad 211 002, INDIA\\
$^*\/$ Also at Jawaharlal Nehru Centre for Advanced
Scientific Research, Bangalore 560 064, INDIA\\
$^{\S} $ Partly supported by IFCPAR.\\

\newpage

\noindent 1. {\bf Introduction}\\

The discovery of high temperature superconductors has
led to a renewed interest in the mixed phase.  Several
novel phenomena arising from their short coherence
length, layered nature and large superconducting
transition temperatures have been theoretically and
experimentally studied(1).  One area of interest is
the existence of effects connected with vortex
dynamics, e.g. quantum creep (2,3), anomalies in the
Hall effect (4-6), and ac electromagnetic response
(7,8).  These phenomena are not fully understood,
partly because of the lack of a well developed first
principles theory of vortex dynamics, especially in
the quantum, interacting vortex regime.  A number of
recent contributions address parts of the problem
(9-12), especially the Magnus force driven dynamics in
the presence of dissipation.

A necessary ingredient in all considerations of the
motion of a vortex is its inertial mass.  This
quantity, generally believed to be small, is
surprisingly ill known and its origin is not well
understood.  (See Ref. 1,  for example).  Not
much attention has been paid to this question because,
for some phenomena, the dynamics is governed by the
large dissipation (13) or the strong Magnus force (9-12)
and the inertial mass could be irrelevant.  However,
there is experimental evidence for a low dissipation
regime in cuprate superconductors (5), and for a
Magnus force smaller (14) than standard estimates
(9-12).  It is thus quite possible that the inertial
mass could affect dynamical processes involving
vortices. (These questions are taken up in Section
IV).  Also, in the absence of an understanding of what
contributes to the vortex mass, and how much, it is
difficult to meaningfully discuss the question of
whether or how such a mass influences vortex dynamics.
 We therefore present here an extensive discussion and
a calculation of the vortex mass, and go into the
question of mass related phenomena in Section IV.  This
work was first reported in 1994(15).

We first estimate the mass using a phase-only 
functional (Section II) which is known to give a good description of the 
system far from the vortex core, where the amplitude of the order parameter
is nearly constant and the only relevant degree of freedom is the phase.
 Recently, Duan and Leggett (17)
and Duan (18) have given a careful discussion of this
approach where the time dependent order parameter (phase) of a moving
vortex causes
 the electronic density to fluctuate. This in turn
gives rise to an electric field which is
screened. The  energy of the density distortion and 
electric field energy are the cause of the mass.
The phase-only functional used has been derived microscopically (19)
and the result obtained thus has a microscopic 
significance and gives an accurate estimate of
the contribution to the vortex mass which accrues from
transitions induced in the electronic scattering states
of a vortex by the vortex motion.
The contribution of this process from the far region
turns out to be very small due to the efficient screening; thus 
most of the mass comes from the core of the vortex.
 
We then calculate the core contribution to the 
mass using a phenomenological Ginzburg-Landau
functional   as has been conventional since the early
work of Suhl (16). This functional is not derivable
microscopically and is used mainly as an interpolating
formula which reduces to the correct phase-only functional in the far
region. However, obviously the coarse grained GL approach is
unrealistic for effects within the core which is of the same size
($\xi$) as the coarse graining scale of the theory.
Further the screening length in GL theory is proportional
to $|\psi|^{-1}$, where $\psi$ is the superconducting order
parameter, and thus diverges at the core.
This is clearly an artefact of the GL approach, as the
efficiency of Coulomb screening is related to the
electronic compressibility, and is expected to be
largely independent of the superconducting order
parameter.  
The GL estimate of the vortex mass is clearly unreliable
and is presented here mainly to contrast the correct microscopic
calculation of the mass which forms the main body of this paper.
The correct, microscopically obtained vortex mass,
is very different from the GL
estimate in its dependence on the basic parameters
of a superconductor.
Inspite of this, the numerical value of the mass 
obtained in the GL approach 
is (largely by accident) in the same range as that obtained from the
microscopic theory for
parameters appropriate to the cuprates.

We then present the first correct microscopic calculation  of the vortex
mass (Section III). We employ the self consistent pair field
approximation
 which has been used extensively for
static vortex structure, quasiparticle energy levels,
etc. (20-23).  We make a Galilean transformation to
a frame of reference where the vortex is at rest.  In
this frame, the motion of the vortex acts like a
perturbation of the form $\vec{u}.\vec{p}^{op}\/$
where $\vec{u}\/$ is the vortex velocity and
$\vec{p}^{op}\/$ is the momentum operator for the
electrons.  The inertial mass is obtained by
integrating out the electronic degrees of freedom to
second order in $\vec{u}\/$.  The coefficient of the
($u^2/2\/$) term in the effective action is
the effective mass of the vortex. The
mass is found to originate from a polarization process
involving the virtual excitation of the lowest energy  quasiparticles in
 the bound electronic
states (quasiparticle states) localized in the core of
the vortex.   The small core size ($\xi \sim 15
\AA\/$) in the cuprate superconductors implies that
the lowest unoccupied state is separated by a sizeable
gap ($ \sim 100 K\/$) from the highest occupied
state below the Fermi level (23), in strong contrast
to conventional superconductors. 
The existence of this large gap in the core quasiparticle spectrum
has been recently observed (43) by STM measurements in
 the vortex core region
in $YBa_2Cu_3O_{7-\delta}$.
This virtual
transition process gives
rise to a large vortex mass ($m^* \approx 25 m_e\/$).
However, strong dielectric screening drastically
reduces the mass and leads to a value $m^* \simeq
0.5 m_e\/$ per $CuO_2\/$ layer.  We discuss the
physical reason for a mass of this size in terms of
the basic length scales and the screening process.

In the final section (Section IV) we discuss the calculation
critically, compare with other results, consider the calculation of a
vortex mass for a non s-wave superconductor and go into the 
question of when effects due to the small vortex mass
might be observable.\\

\noindent II. {\bf Ginzburg Landau calculation of the vortex
effective mass}

The most natural way of discussing the motion of a
singularity in the phase $\theta\/$ of the
superconducting order parameter is the time
dependent Ginzburg Landau theory (17 - 19) where the
free energy (or action) is expressed as a functional
of the phase of the 
superconducting order parameter.
This functional provides a good description
of the region far from the centre of the vortex
where the amplitude of the superconducting gap is nearly constant.
  We describe the functional and briefly
summarize  known results for the mass contribution
from the region outside the core (17,18).

Outside the core, a phase only Hamiltonian is
sufficient, and the action functional {\it S} per 
length {\it L} (at $T=0$) is given by 
$$(S/L) = S_{\theta} + S_{em} \eqno{(1a)}$$
where
$$
S_{\theta} = \int^{+\infty}_{-\infty} dt \int d \vec{r}
\left[ \frac{\alpha_1}{2} \left( \dot{\theta} -
\frac{2eA_0}{\hbar} \right)^2 - \frac{\alpha_2}{2}
\left(({\nabla}\theta - \frac{2e\vec{A}}{\hbar
c})^2 +{4m\dot{\theta}\over\hbar}\right)\right] \eqno{(1b)}$$
and 
$$S_{em} = \int^{+\infty}_{-\infty} 
dt   \int d\vec{r} \left[ \frac{(\nabla A_0)^2 -
({\nabla} \times \vec{A})^2}{8 \pi} \right] \eqno{(1c)}$$
where $\vec{r}$ is the coordinate in the plane
perpendicular to the magnetic field.  In Eq. (1b), the first two
terms are the energies associated with pair charge and
pair current (or velocity) fluctuations respectively.
The term linear in $\dot{\theta}$  is a total time derivative
which has no physical consequences in the absence of vortices.
Its physical origin can be understood (10) in terms of the Berry phase 
associated with the adiabatic motion of a vortex and it gives rise to the
Magnus force. Notice the absence of any term linear in $A_{0}$ which
is  constrained to be zero because of charge neutrality in 
the electron-ion system.

The action functional of Eq. (1a) can be obtained microscopically
by starting with electrons interacting with a pair potential
(whose magnitude is nearly constant in the far region),
going to a gauge where the order parameter is real
and then integrating out the fermions. The details of the derivation
have been outlined in (19). In Fig. (1), we show the Feynman diagrams
which contribute to the action of Eq. (1a) at $T=0$ in the clean limit.
The coefficients
$\alpha_1\/$ and $\alpha_2\/$ are the appropriate
polarizabilities. For a weakly
interacting, clean Fermi gas, they have the values
$$\alpha_1 = \left(\frac{\hbar}{2e}\right)^2
\frac{\lambda^{*-2}_{TF}}{4 \pi} \eqno{(2a)}$$
and 
$$ \alpha_2 = \left(\frac{\hbar c}{2e} \right)^2 \frac{\lambda^{-2}_L}{4
\pi} \eqno{(2b)}$$
where $\lambda^*_{TF}$ (the Thomas Fermi screening
length) is given by 
$\lambda_{TF}^{*-2} =  (6\pi ne^2/\epsilon_{F})$ and 
$\lambda_L\/$ (the London penetration depth) is given by
$\lambda^{-2}_{L} = (4 \pi n e^2)/(m_e c^2)\/$.  
 $A_0\/$
and $\vec{A}\/$ are the scalar and vector electromagnetic
potentials respectively, the two terms in Eq. (1c)
being just the field energies.  In case there is an
external magnetic field $\vec{H}_0\/$, the term
$\left\{ ({\nabla} \times \vec{A})^2/8 \pi
\right\}\/$ is modified to $\{({\nabla} \times \vec{A} - 
\vec{H}_0)^2/8 \pi\/\}$.
 The functional is clearly gauge invariant. 
  In Eq. (1) the
displacement current term $({1\over c}{\partial \vec{A}\over
\partial t}  )^2\/$ has
been omitted because of the largeness of c.

The motion of the vortex in a charged superconductor 
gives rise to an electric
field (or potential $A_0\/$).
To find this $A_0\/$, we minimize $S\/$ with
respect to $A_0\/$ which yields
$${\nabla^2 A_0\over 4\pi} =  \frac{-2e}{\hbar}
\alpha_1 \left(\dot{\theta} - \frac{2eA_0}{\hbar}\right)
 \eqno{(3)}$$
Now for a vortex moving with a (small) uniform
velocity $\vec{u}\/$, we make the assumption that 
$$\theta(\vec{r},t) = \theta_0(\vec{r} - \vec{u}t) \eqno{(4)}$$
where $\theta_0(\vec{r})\/$ is the phase around a
static vortex at the origin. Thus we have 
$$
\dot{\theta} (\vec{r}, t) = -\vec{u}. \vec{\nabla} \theta_0 
\eqno{(5)}
$$
Using this in combination with the known $\theta_0\/$,
one finds the potential $A_0\/$ from Eq. (3) and the
extra energy from Eq. (1).  The result is 
$$E_{KE}={\pi\alpha_1u^2\over 4}ln(1+{{\xi^{-2}-R_c^{-2}}\over{\lambda_{TF}
^{*-2}+R_c^
{-2}}})\eqno{(6a)}$$
where $R_{c}$ is a long distance cutoff (whose magnitude will
be taken to infinity in the end). 
Then it is easy to see that in the limit $e\rightarrow 0$
which implies $\lambda_{TF}^{*}\rightarrow \infty$, the expression for 
$E_{KE}$ reduces to
${{\pi\alpha_1u^2}\over{2}}\,ln({R_c\over \xi})$
which diverges logarithmically as the long-distance cutoff $R_c$
is taken to $\infty$. This is the expected result for a neutral superfluid
(17, 18). For the present case of a charged superconductor
we are always in the limit $\lambda_{TF}^*\ll\xi$ and the
expression for the kinetic energy simplifies to
$$E_{KE}\simeq {{\pi\alpha_1u^2}\over 4}{\lambda_{TF}^{*2}\over \xi^2}=
{u^2\over 16\xi^2}\left ({\hbar\over 2e
}\right )^2\eqno{(6b)}$$
The expression in Eq. (6b) can be rewritten as 
$$
E_{KE}=  E_{SV} \left( \frac{u^2}{v^2_F}\right)
\left(\frac{\lambda^{*2}_{TF}}{\xi^2}\right) \left(
\frac{3}{4ln(\lambda_L/\xi)} \right) 
\eqno{(6c)}
$$
where $E_{SV}=({\phi_0\over 4\pi\lambda_L})^2\,ln({\lambda_L\over \xi})\/$ 
is the London energy (per unit length) of a static vortex
which results from the terms involving transverse fluctuations
in the current and magnetic field in the phase functional of
Eq. (1).  The result
is then easily understood on physical grounds; the velocity
$u\/$ is to be compared with the natural velocity
scale $v_F\/$ of the Fermi system.  The second
bracketed factor in Eq. (6c) is due to the reduction of charge
fluctuations by
screening.  The former occur on a length scale
$\xi\/$, while the screening length is $\lambda_{TF}^*
\ll \xi\/$.  On substituting appropriate numbers in Eq. (6b), i.e.
$\xi \simeq 15 \AA\/$  and $\lambda_{TF}^* \simeq 1 \AA\/$
 we find
that the contribution of this source to $m^*_f \simeq 7
\times 10^{-4} m_e\/$ per layer, ($m_e$ being the electron mass) 
assuming an interlayer spacing $ d \simeq 10
\AA\/$. This is an extremely small
number; one reason for its smallness is the
screening factor $(\lambda_{TF}^*/\xi)^2 \simeq (1/200)$.
The result,  (Eq. (6b)),  has been obtained earlier by Duan
(18), where details may be found.

The functional $S_{\theta}$ of Eq. (1) is not Galilean invariant
and there are additional terms involving  higher order 
derivatives of $\theta$,
 whose inclusion is necessary to restore this symmetry.
These arise from the fact that the Gaililean invariant combination involving
the time dependent order parameter phase $\theta$ is the local
electrochemical potential
$\delta\mu({\vec r},t)=-{\hbar\over 2}{\dot{\theta}}+eA_0 -
{({\hbar\nabla\theta\over 2}-{e{\vec A}\over c})^2\over 2m}$
The correct functional was recently obtained by Aitchison and co-workers
(24) and is of the form 
$$S_{\theta}=\int^{+\infty}_{-\infty} dt \int d \vec{r}
 \left[{\alpha_1\over 2}[({\dot{\theta}}-{2eA_0\over \hbar}) +
{\hbar(\nabla
\theta-{2e{\vec A}\over \hbar c})^2\over 4m}]^2-
{\alpha_2\over 2}[{4m\over\hbar}{\dot{\theta}} +
(\nabla
\theta-{2e{\vec A}\over \hbar c})^2]\right]\eqno(7a)$$
The additional diagrams contributing here, are shown in Fig. (2).
The harmonic electric field terms in Eqs. (7a) and (1c)
can be integrated out to give the action functional
$$S_{\theta}=\int dt\int {d\vec{q}\over (2\pi)^2}{2\alpha_1\over \hbar^2}
\mid\delta\mu({\vec q}){\mid}^2 {q^2\over {q^2+\lambda_{TF}^{-2}}}
-{\alpha_2\over 2}\int dt\int d{\vec r}[{4m\over\hbar}{\dot{\theta}} +
(\nabla
\theta-{2e{\vec A}\over \hbar c})^2]\eqno(7b)$$
It is clear from Eq. (7b), that additional contributions accruing to the
vortex mass come from the coupling of charge fluctuations
to the supercurrent fluctuations leading to non-adiabatic
corrections in the  supercurrent distribution that are
proportional to the vortex velocity.
 However, an explicit calculation shows
that all such contributions to the vortex mass are screened
at least doubly more efficiently and are therefore 
smaller than that estimated
above (Eq. (6b)) by a factor of $(\lambda_{TF}/\xi)^2\approx .005$ .
Since the far mass estimated before  is already small,
this additional contribution is smaller still, and
therefore negligible. In fact in the perfect screening approximation,
where the total electro-chemical potential $\delta\mu$ is set
equal to zero locally, we get back the result of Eq. (6), the
contribution of the additional terms being zero as expected.
It is worth emphasizing that the results obtained from the phase-only
functional are essentially microscopic (see the Feynman diagrams in Fig. 1
and Fig. 2) and give an accurate estimate of the contribution to the vortex
mass from the electronic scattering states which are extended in nature
and live mainly
outside the core.

We now consider the contribution to the vortex  mass
arising from the core.  The amplitude of the superconducting 
order parameter changes as a function of radial distance
from the center of the vortex
in the core region. We use 
a phenomenological Ginzburg - Landau action functional per unit length
$$S  =  \int dt \int d\vec{r} \left[ \frac{3}{2
mv^2_F} \left| \left( \frac{\hbar}{i} \frac{\partial}{\partial
t} - 2eA_0 \right) \psi (\vec{r},t) \right|^2 - \frac{1}{2m} \left|
\left(\frac{\hbar}{i} {\nabla} -
\frac{2e}{c}\vec{A} \right) \psi \right|^2
- V(|\psi|^2) \right]$$
$$ + \int dt \int d\vec{r} \left[ \frac{({\nabla} {A_0})^2
- ({{\nabla} \times \vec{A})^2}}{8 \pi}\right]\nonumber \eqno{(8)}$$
to estimate the core contribution to the vortex mass.

This functional is not derivable microscopically
and has been chosen primarily as an interpolating formula
which reduces to the correct phase-only functional
in the far region in the limit $\Psi\rightarrow \sqrt{n_s}e^{i\theta}$
and gives the correct (linear in $r$) dependence for the amplitude of
the order
parameter near the center of the vortex. 
The  parameter $m$ which appears above
is therefore fixed by requiring this functional
to reduce to the action functional of Eq. (1) in
the ``phase-only" approximation. For the potential
$V(|\psi|^2)$ we assume the standard form
$V(|\psi|^2)={\alpha\over 2}|\psi|^2+{\beta\over 4}|\psi|^4$.

Starting with the early work
of Suhl (16) and subsequent work by others (see Ref. (1) and references 
therein)
 all estimates of the vortex
mass have proceeded from this functional. The value of the mass 
obtained from this functional is not expected to be very accurate for
reasons that are discussed below. However we present a calculation of the
vortex mass from this functional mainly to contrast and highlight the
microscopic calculation presented in Sec. III.

We again (see Eq. (4)) make the ansatz that for small
velocity $\vec{u}\/$, the vortex moves rigidly i.e.
$$\psi(\vec{r},t) = \psi_0(\vec{r} -\vec{u}t)\eqno{(9)}$$
where $\psi_0(\vec{r})\/$ is the order parameter
configuration associated with a static vortex.  We
further assume a common and fairly accurate explicit
form for $\psi_0(\vec{r})\/$, namely
$$\psi_0(\vec{r}) = \sqrt{n_s} \tanh (r/\xi) \exp(i \phi) \eqno{(10)}$$
where $n_s\/$ is the superfluid density far from the
vortex, and $(r, \phi)\/$ are the radial and angular
coordinates of the two dimensional vector $\vec{r}\/$.
 Again, minimizing Eq. (7) with respect to $A_0\/$, we find
 $${\nabla^2 A_0\over 4\pi} =  \frac{3 e \hbar}{imv^2_F}
\left( \psi \frac{\partial \psi^*}{\partial t} - \psi^*
\frac{\partial \psi}{\partial t} \right) +
\frac{3}{mv^2_F} (2e)^2|\psi|^2 A_0  \eqno{(11)}$$.
Using this condition in the action functional of 
Eq. (8) we find that the  extra energy  due to vortex motion
is

$$ \Delta E = \int d\vec{r} \left[
\frac{3 \hbar^2}{2mv^2_F} |\frac{\partial \psi}{\partial
t} |^2 - \frac{3e\hbar A_0}{2mv^2_Fi} \left( \psi^*
\frac{\partial \psi}{\partial t} - \psi \frac{\partial
\psi^*}{\partial t} \right) \right] \eqno{(12)}.$$

The first term in Eq. (12) is due to the density
fluctuations induced by vortex motion, and the second describes the
reduction  due to screening.  The first term
is readily computed using the ansatz Eq. (9) for the
time dependence of $\psi({\vec r},t)$ and the form Eq. (10) for the
coordinate dependence of $\psi_0(\vec{r})\/$.  
We are interested here only in the core contribution to the
mass, the contribution from the far region having been previously
determined (Eq. (6)).
The radial integration in Eq. (12) is therefore performed
over the range $0<r<\xi$ (core region).
It
gives a mass per unit length
$$
m^{* \;\;\;\;\;\;\;\; \mbox{core}}_{\;\;\mbox{unscreened}} 
\simeq m^{*\;\;c}_{ u}  =0.61
\left( \frac{m_e}{a_0} \right) \left(
\frac{a_0}{4 \lambda^*_{TF}}\right)^2\eqno{(13)}$$
where $a_0={\hbar^2\over m_e e^2}\/$ is the Bohr radius, and
$(\lambda^*_{TF})^{-2} = 4 \pi n_s (2e)^2 /(mv^2_F/3)\/$.
 For the cuprate superconductors,
we find
that the unscreened core vortex mass per layer is 
$$m^{*\;\;c}_{ u} \approx 0.19 m_e \eqno{(14)}$$
An unexpected, and incorrect, feature of Eq. (13) is the
lack of dependence of the vortex mass on the core size $\xi\/$.
This is a consequence of the fact that the core mass
is proportional to the 
 gradient energy per unit length in the core,
which is scale invariant and does not depend on the natural length
scale $\xi\/$ in two dimensions. This is inevitable in any
local continuum free energy theory.

We now consider the second or screening term in
Eq. (12).  One clearly needs  to know the electric
potential $A_0\/$ induced by vortex motion.  Setting 
$$A_0(\vec{r}) = v(r) \vec{u}\cdot \hat{\phi}\eqno{(15)}$$
we find that $v(r)\/$ satisfies the radial equation
$$\left[ \frac{\partial^2}{\partial r^2} + \frac{1}{r}
\frac{\partial}{\partial r} - \frac{1}{r^2} -
\frac{1}{(\lambda^*_{TF})^2} \frac{|\psi(r)|^2}{n_s}
\right] v(r)$$
$$ = \frac{1}{4 \pi} \frac{\hbar}{2e}
\lambda^{*-2}_{TF} \left[ \frac{\tanh^2
(r/\xi)}{r} \right] \eqno{(16)}$$
This is just the  radial Poisson equation
with a screening term (last term on the left hand
side), and a source term (on the right hand side) appropriate for
a two dimensional system.
The `effective screening length' $\lambda^*_{TF} \sqrt{n_S}/|\psi(r)|\/$
 is $r$ dependent, and increases
as $|\psi(r)|\/$ decreases, i.e. with $r \rightarrow
0\/$.  This is yet another unrealistic feature of the
Ginzburg Landau theory, since one expects screening 
which is related to the electronic compressibility, to be
relatively independent of superconducting order.
Eq. (16) is solved numerically, with appropriate 
boundary conditions at $ r=\infty\/$ and $r=0\/$. The
$v(r)\/$ and the $A_0(\vec{r})\/$ thus obtained (see
Eq. (15)) are used to calculate the second term in the
vortex core energy (Eq. (12)).  Adding the contributions
of both the terms, the final result for the mass per layer
is 
$$m^*\;\;\; \mbox{(core)} = m_c^* =\left( \frac{a_0}{4
\lambda^*_{TF}} \right)^2 \left(\frac{d}{a_0}\right)
\left[0.61 - 0.04 \right] m_e \eqno{(17)}$$
where $d$ is the interlayer separation.  In Eq. (17),
the first term in the square brackets is due to the
density distortion in the core, and the second term
is the negligibly small correction due to screening.  Thus the total
mass per layer, for a cuprate superconductor at $T =
0 K \/$ is 
$$m^*_c  \simeq 0.19 m_e \eqno{(18)}.$$
In the absence of a microscopic theory, the phenomenological 
Ginzburg Landau approach has been used, mainly as a dimensional
aid, to estimate the vortex inertial mass.
The values quoted 
( Ref. (1) )
are in
the range 0.2 to 2 $m_e\/$, and turn out to be (largely by accident) 
not far
from our microscopic result (see below) for parameters
appropriate to the cuprates.
As a preliminary to the microscopic calculation which
forms the main result of this paper, we have performed a
detailed calculation using this functional to bring out its
inadequacies.
  The above
detailed analysis of the different  contributions to the mass shows that
within the phenomenological GL theory it is due to 
the essentially unscreened density distortion
induced in the core. 
The unscreened
mass estimated above too is incorrect as
the gradient expansion implicit in a phenomenological
theory like the GL functional of Eq. (8), breaks down
at short length scales and strongly non-local (in space)
effects lead to a much larger mass (see Sec. III. below) 
 in the unscreened
case than calculated here.
This picture is clearly incorrect on another count as well.
Electronic screening processes are not expected to be affected much
by the onset of superfluid order and a strong reduction of the mass
is expected because of Coulomb screening. 
This aspect too is explicitly seen in the microscopic theory.
In short, the microscopic calculation shows that the phenomenological
GL picture is wrong on all counts, as we shall see now.

\noindent {\bf III. Microscopic calculation of the vortex mass}

We present now a microscopic calculation of the vortex inertial
mass for a layered superconductor at $T=0\/$.  The
dynamics of the system of paired
electrons is described by the action

$$S = \int dt \int dz \int d\vec{r} \sum_l \left[
L_l^f (\vec{r} ,t) + {\tilde L}_l^f(\vec{r},t)+
L^{\mbox{\it pair}}_l (\vec{r}, t)
\right] \delta(z-ld) + S_{em}  \eqno{(19a)}$$
where
$$L^f_l (\vec{r}, t) = \sum_{\sigma} \bar{\psi}_{l
\sigma} (\vec{r}, t) \left[ i \hbar
\frac{\partial}{\partial t}  
-{(\frac{\hbar}{i} {\nabla} - \frac{e }{c}\vec{A})^2\over 2m}
 + \epsilon_F \right] \psi_{l \sigma} (\vec{r},t) \eqno{(19b)}$$
$${\tilde L}^f_l({\vec r},t)=-eA_0({\vec r},ld)
\sum_{\sigma} \bar{\psi}_{l\sigma}
(\vec{r},t)\psi_{l\sigma}(\vec{r},t) \eqno{(19c)}$$
$$L^{\mbox{\it pair}}_l (\vec{r}, t) = 
-[\Delta_l(\vec{r}, t) \bar{\psi}_{l\uparrow}(\vec{r}, t)
\bar{\psi}_{l\downarrow} (\vec{r},t) + h.c. ] -
{|\Delta_l({\vec r},t)|^2\over V}
\eqno{(19d)}$$
and
$$S_{em}=\int dt\int dz\int d\vec{r}{({\nabla A_0})^2-({\nabla\times{\vec
A}})^2\over 8\pi}\eqno(19e)$$

The electrons at $(\vec{r}, t)\/$ on layer $l$, with
spin $\sigma\/$ are represented by the Grassmann field
variables $\bar{\psi}_{l \sigma}(\vec{r},t), \;\;
\psi_{l \sigma} (\vec{r}, t)\/$. The electronic kinetic energy
is described 
 by the term
$L^f_l (\vec{r}, t), \/$ i.e. Eq. (19b)
and the electronic coupling to the Coulomb potential by the 
term ${\tilde L}^f_{l}(\vec{r},t)$ (Eq. (19c)). 
Eq. (19d) is
the 
mean pair
field decomposition of the Cooper pair attraction,
appropriate for a BCS s-wave superconductor and $V$
is the strength of the attractive contact interaction.  
The last
term $S_{em}$ (Eq. (19e)) is the electro-magnetic field contribution.

Several possible modifications and generalizations,
such as order parameter symmetries other than s-wave and
 effects due to quasiparticles at $T\neq 0$ are briefly
discussed in Section IV.  We have neglected above the
small interlayer (Josephson) coupling, so that the
(pancake) vortices in different layers 

are coupled only via electric and
magnetic fields.  This neglect has very little effect
on the vortex mass, which is overwhelmingly due to
processes occuring within a layer, and as a matter of
fact, to processes within the small core.

Every layer  has one pancake vortex. 
All the pancake vortices are assumed to
lie along a straight line parallel to the magnetic field
which is perpendicular to the layers and to
move with a uniform velocity $\vec{u}\/$, so
that the vortex (core) coordinate is 
$(\vec{R_0}, ld)\/$ where $\vec{R}_0 = \vec{u}t\/$ is a vector 
in a plane parallel to the layers and the $z$ co-ordinate
$ld$ specifies the position of the layer.
The pair potential $\Delta_l(\vec{r}, \; t)\/$ in the
presence of such a uniformly moving vortex is a
function of $ (\vec{r} - \vec{R}_0 (t))
\/$.  It can thus be written in an adiabatic approximation as 
$$
\Delta_{l} (\vec{r}, t) =\Delta_0 (|\vec{r} -
\vec{R}_0 (t)|)\; e^{i \theta (\vec{r} -
\vec{R_0}(t)) }  \eqno{(20)}.$$
Here $\Delta_0(r)\/$ is the magnitude of
the pair potential and $\theta(\vec{r})\/$ the phase,
for a static vortex situated at 
$\vec{r} = 0\/$.  The pair potential does not depend
on $l\/$.  The action functional relevant for vortex dynamics is found by
an expansion of the microscopic action in powers of the vortex velocity $u$
after integrating out the electronic degrees of freedom. The vortex mass
is then determined from the term quadratic in $u$ (the dissipative and
Magnus forces coming from the linear term). This calculation is most
conveniently carried out in the rest frame of the vortex. 
In this frame, the pair potential seen by the electrons is
$\Delta_0 (r)\, 
\exp[i (\theta (\vec{r})
-{2m{\vec u}\cdot{\vec r}\over \hbar}+{mu^2t\over\hbar})]    $
i.e. the pair potential corresponding to a vortex at rest with
 extra phase factors coming from the macroscopic supercurrent
 and kinetic energy of
the electrons which acquire an extra velocity $-{\vec u}$ in this frame.
The Coulomb potential seen by the electrons in this frame becomes
$A_0-{{\vec u}\over c}\cdot {\vec A}$ while the
vector potential is the same as before. All physical
quantities can be calculated 
in this frame and then transformed  to the lab frame as necessary.
Note that this does not assume Galilean invariance.

After a gauge transformation, the action for the system in this moving
frame can be written as
$$ S = S_0 + S_1 \eqno{(21)}$$
where the static vortex action $S_0$ is 
$$
S_0 = \int dt \int d\vec{r} \sum_l \left[
L^f_l (\vec{r}, t)- \Delta_0(r)[e^{i
\theta (\vec{r})} \bar{\psi}_{l \uparrow} (\vec{r}, t)
\bar{\psi}_{l \downarrow} (\vec{r}, t) + h.c.] -
{\Delta_0^2(r)\over V}\right]
$$
$$
- \int dt \int dz \int d\vec{r}  {({\nabla}
\times \vec{A})^2 \over 8 \pi}  \eqno{(22a)}
$$
and the perturbation due to vortex motion is contained
in 
$$S_1 = \int dt \int d \vec{r}  \sum_l \left[ 
\vec{u} \cdot \vec{p}_l (\vec{r}, t) 
-e A_0(\vec{r},ld) \rho_l(\vec{r}, t)\right]
 + \int dt \int dz \int
d\vec{r}  {(\nabla A_0)^2\over 8 \pi}  \eqno{(22b)}$$
In Eq. (22b), $\vec{p}_l (\vec{r}, t)$ is the momentum
density operator for the $l\/$th layer,
$$\vec{p}_l (\vec{r}, t) = \sum_{\sigma} \bar{\psi}_{l
\sigma} (\vec{r}, t) {\hbar\over i} {\nabla}
\psi_{l \sigma} (\vec{r}, t) \eqno{(23a)} $$
and  $\rho_{l}(\vec{r}, t) \/$ is the density operator
$$\rho_{l}(\vec{r}, t) = \sum_{\sigma} \bar{\psi}_{l
\sigma} (\vec{r}, t) \psi_{l \sigma} (\vec{r}, t) \eqno{(23b)}$$
The first term in Eq. (22b) is linear in $\vec{u}$,
coupling to the electron momentum.  The second term is
the electric potential energy of the (nonuniform) electron density
around the moving  vortex.  The associated electric potential
 has to be determined self consistently.

Since the 
  London screening
  length ($\lambda$) is much larger than the core size ($\xi$), the magnetic
  field in the core is nearly the same as the external magnetic field,
deviations from this being of order $(\xi/\lambda)^2$.
  However, the vector potential associated with this field 
  is negligible in comparison with the gradient of the phase
  of the superconducting order parameter and has been ignored in the
  following calculations.
  The former $\approx H\,r$ whereas the latter $\approx \phi_0/r$.
  Thus for the core region ($r\leq\xi$), we find that for $H\ll \phi_0/\xi^2 $
  the vector potential may be ignored (20).
  Corrections due to the vector potential can be estimated to
  be of order $H/H_{c2}$ where $H$ is the external magnetic field
  and $H_{c2}$ is the upper critical field and are therefore small
  in the dilute vortex limit.

Now a systematic expansion in powers of ${\vec u}$
becomes possible by integrating out the electrons 
giving rise to the Lagrangian
that describes vortex dynamics.
Dynamics of vortices (classical or quantum) 
 can be studied either
by working directly with the vortex action or alternately by
introducing a canonically conjugate momentum
which permits one to go over to the Hamiltonian
formalism.

The Magnus and dissipative forces come from
the term linear in ${\vec u}$ and can be obtained
by taking the gradient of this term with respect to the
vortex co-ordinate (37).  The inertial term in the action
is quadratic in the vortex velocity.
We are thus interested in the change of action to second
order in $\vec{u}\/$, the coefficient of ($u^2\/$/2)
in the change being the vortex effective mass $m^*\/$.
Clearly, this is calculable by going to second order
in $u$ and $A_0$.  For ease of
presentation
and also to emphasize the importance of Coulomb screening
in the core, we do this in two stages.  First, we
find the unscreened vortex mass, i.e. in $S_1\/$ (Eq.
(22 b)), we turn off the Coulomb interaction by
putting $e=0\/$, and calculate the second order shift.
 We then calculate the effect of Coulomb interactions,
i.e. the effect of the electric potential due to the
electron charge density change consequent on vortex motion.
We would like to emphasize that the unscreened mass calculated here
corresponds to the contribution of the {\it {bound states}} localised
in the vortex core. 
The contribution of bound states is finite.
For a (hypothetical) neutral superconductor,
this term is to be added to the contribution
from the region outside the core.
This can be calculated from the (microscopically
derived) phase-only functional (Eq. 1) by
setting $e=0$, and as mentioned earlier,
is logarithmically divergent because 
longitudinal density fluctuations
associated with vortex motion are not
screened. Thus for the uncharged superconductor,
the finite core contribution can be neglected
in comparision with the log diverging far contribution.

For a charged superconductor, the core contribution 
has to be calculated with the inclusion of 
screening effects which, we show ( Eq. (43) below)
reduces the `unscreened' or $e=0$ mass by a factor
of fifty or more.
To this we have to add the $m^*$ due to the far region,
which because of Coulomb screening is finite, and
is actually negligibly small (Eq. (6c)).

The second order effective Lagrangian due to just the
$\vec{u}. \vec{p}\/$ term in $S_1\/$ is given using
standard many body perturbation theory by $ {m^*_0\over 2}
u^2\/$ where 
$$m^*_0 = i\;\; \int  d\vec{r} \int
d\vec{r'} \int dt \left<T\left[p^x_l(\vec{r},t)
p^x_l(\vec{r'}, 0)\right] \right> 
=-\int d\vec{r}\int d\vec{r'}
\chi^{xx}_l({\vec {r}},{\vec{r'}}) \eqno{(24)}$$
In Eq. (24), the vortex velocity has been assumed to be
in the x-direction. 
Note that intra-layer averages such as Eq. (24)
 do not depend on the layer index.  The
correlation function in Eq. (24) is calculated with
respect to the unperturbed action $S_0\/$ (Eq. (22a))
which describes a single static vortex.  This last problem
of a static vortex 
has been studied extensively using the
Bogoliubov-de Gennes self-consistent field theory
(20-23).  The electronic eigenstates, in the presence
of a static vortex, are 
 bound states which are localised and have appreciable amplitude
only in the core of the vortex, or extended states
which are scattered by the superfluid velocity
and are primarily in the region outside the vortex core
where the amplitude of the order parameter is nearly
constant. The latter are well described by
neglecting the variation in the amplitude of the order
parameter and the scattering processes contributing to the mass
involving these states are just those considered in the Feynman 
diagrams (Fig. 1) contributing to 
the phase-only functional in Eq. (1).
Thus the mass contribution from the deformation of the scattering
states is accurately estimated by the calculation proceeding
from the phase-only functional of Eq. (1) and has
been shown to be negligibly small due to efficient screening.
We will therefore concentrate in the following only
on the contribution of the localised states to the correlation function
of Eq. (24) to find the core contribution to the vortex mass.

The eigenfunctions of the Bogoliubov-de Gennes
 equations for a single layer which are localised in the vortex core 
 are the amplitudes
$u_{\mu}(\vec{r})\/$ and $v_{\mu}(\vec{r})\/$,
labelled by the (azimuthal) angular momentum quantum
number $\mu\/$ because of the cylindrical symmetry of
the single vortex problem. 
The Bogulibov amplitudes are related to the fermion field
operators by the relations 
$$\psi_{\uparrow}(\vec{r}, t) = \sum_{\mbox{all} \mu}
u_\mu (\vec{r}) \gamma_\mu(t) \eqno{(25a)}$$
and 
$$\psi_{\downarrow}^{\dagger}(\vec{r}, t) = \sum_{\mbox{all} \mu}
v_\mu (\vec{r} ) \gamma_\mu(t) \eqno{(25b)}$$
Here $\mu$ runs over all half-odd integers (positive
as well as negative). The quasi-particle annihilation 
operators $\gamma_{\mu}$ correspond to the empty (particle)
states for $\mu >0$ and filled (hole) states for $\mu <0$.
In terms of the quasi-particle operators $\gamma_{\mu}$,
the Hamiltonian for the static vortex system described by the
action $S_0$ can be written in a diagonal form 
as,
$$H_0 = \sum_{\left\{\mu\right\}} \epsilon_\mu \gamma^+_\mu
\gamma_\mu + \;\; \mbox{constant} \eqno{(26a)}$$
Our definition of the quasi-particle operators (Eq. (25))
differs from the standard definition (20) by a particle-hole
transformation. In doing this, use has been made of the fact
that if $(u_{\mu}({\vec r}),v_{\mu}({\vec r}))$ is an eigenstate
with eigenvalue $\epsilon_{\mu}$, the state $(-v_{\mu}^*({\vec r}),u_{\mu}^* 
({\vec r}))$ is also an eigenstate with eigenvalue $-\epsilon_{\mu}$.
Thus, 
$$(u_{\mu}({\vec r}),v_{\mu}({\vec r}))=(-v_{-\mu}^*({\vec r}),
u_{-\mu}^*({\vec r}))\eqno {(26b)}$$ 
with 
$$\epsilon_\mu = -\epsilon_{-\mu}
\eqno(26c)$$
for $\mu <0$. The ground state of the system has all states
with $\mu <0$ occupied and all states with $\mu >0$ empty.

The correlation function of Eq. (24) is easily
evaluated in this representation. The vortex mass can then be understood
as arising from a
polarization process involving a virtual (quasi-particle) transition
from the highest occupied to the lowest unoccupied state.  This process
can also be viewed as a deformation of the ground state
by the perturbation which mixes in higher energy states.  In
terms of $u\/$ and $v\/$, the momentum-momentum
correlation function  can be written as
$$i \int dt \left< T \left[ p^x_l(\vec{r},t)
p^x_l(\vec{r'}, 0)\right] \right>$$
$$
= 2 \sum^{unocc.}_{\mu>0}
\sum^{occ.}_{\mu'<0} \left( \epsilon_\mu -
\epsilon_{\mu'} \right)^{-1} \times
$$
$$
\left[ u^*_\mu (\vec{r}) {\hbar\over i} {\partial 
u_{\mu'} (\vec{r})\over \partial x}
 \left( u^*_{\mu '}(\vec{r'})
{\hbar\over i} {\partial  u_{\mu} (\vec{r'})\over \partial x'
} - v_\mu
(\vec{r'}) {\hbar\over i}{\partial
 v^*_{\mu '}
(\vec{r'})\over\partial x'} \right) + h.c. \right] \eqno{(27)}.
$$
The various terms here correspond to the different quasi-particle 
processes
that contribute to the polarization.  There is a
simple selection rule for nonvanishing matrix
elements, namely $\mu = -\mu ' = 1/2\/$.  We discuss
this now. The details are worked out in Appendix I.
The operator $p_x$ has an angular dependence of the form $\cos \phi$ (radial
derivative term) and $\sin
\phi\/$ (angular derivative term) 
where $\phi\/$ is the angle in the 2d
plane of the vector $\vec{r}\/$ with respect to the
$x \/$ axis.  The amplitudes $u_{\mu}(\vec{r}), \;
\; v_{\mu} (\vec{r})\/$ can be written (20-23) as
$${u_\mu (\vec{r})\choose v_\mu (\vec{r})}  = e^{-i
\mu \phi} {e^{i \phi/2}\;\;\; f^-_\mu(r)\choose e^{-i
\phi/2}\;\;\; f^+_\mu(r)}\eqno{(28)}$$
where $f^{\pm}_\mu (r)\/$ are functions only of the
radial coordinate $r\/$.  The
$\phi\/$ dependence of $u_\mu
(\vec{r}),\;\;v_\mu(\vec{r})\/$  above implies that on
integration of the matrix elements in Eq. (27) over the
angle $\phi\/$,  the only nonvanishing terms are
$(\mu - \mu') = \pm 1\/$.  We \underline{also} need
 one of the states $\mu$ to be unoccupied and the other $ \mu'$ 
 to be occupied.
 The only possibility among bound states is $\mu =
-\mu'=\frac{1}{2}\/$.  To understand this, we exhibit
in Fig. 3, the spectrum of eigenstates for parameters
appropriate to the  cuprate superconductor (23).
The bound states within the gap have quantum numbers
$\mu = \pm \frac{1}{2}, \; \pm \frac{3}{2},
........\/$.  
The occupied (unoccupied) states have negative
(positive) $\mu\/$.
  It is now clear
that the only virtual transitions that satisfy the
selection rule $\Delta \mu = \pm {1}\/$, \underline{and}
are between occupied and unoccupied
states, correspond to $\mu = -\mu ' =
\frac{1}{2}\/$, i.e. the highest occupied and lowest
unoccupied bound states.  With this simplification used in
Eq. (27) and the latter substituted into the expression
Eq. (24) for $m^*_0\/$, we have
$$m^*_{0} = {4
|g_x|^2\over\epsilon_{1/2}} 
\eqno(29a)
$$
where
$$
g_x = \int d\vec{r} v_{1/2} (\vec{r})
{\hbar\over i}{\partial  u_{1/2}(\vec{r})\over \partial x} \eqno{(29b)}
$$

We thus need to know the energy $\epsilon_{1/2}\/$ and the
core bound state wavefunctions 
 $f^{\pm}_{1/2} (r)\/$
in order to find $m^*_0\/$.  These have been
determined self-consistently and numerically by Zhu
et. al. (22).  We use here the  variational
forms 
$$\Delta_0 (r) = \Delta_0 \tanh (r/\xi) \eqno{(30a)}$$
$$ f^-_{1/2} (r) = A_{1/2}J_0(k_F r) e^{-r/2 \xi} \eqno{(30b)}$$
and
$$ f^+_{1/2} (r) = A_{1/2}J_1(k_F r) e^{-r/2 \xi} \eqno{(30c)}$$
where 
$$A_{1/2}^{-2}=\int d{\vec r} [J_0^2(k_Fr)+J_1^2(k_Fr)]e^{-r/\xi}\eqno
{(30d)} $$
is the normalisation factor.
With $\xi = 15 \AA$, $k_F^{-1}=3.36 \AA$
  and $\Delta_0$ = 60 meV,
the expectation value of the energy
$<\epsilon_{1/2} >\/$ is 69 K, close to the 
self-consistent numerical value of 66 K obtained by Zhu et. al. (22).
The wavefunctions are also very close.  Using these,
the expression $m^*_0\/$ can be evaluated (see Appendix I for details),
 giving a
value 
$$m^*_0 \simeq 25 m_e \eqno{(31)}$$
The vortex mass obtained above can be estimated by the
following simple physical arguments.
The correlation function in Eq. (24) can be estimated as
follows. Each of the momentum operators gives a factor of
$\hbar k_F$ which is the typical electronic momentum.
The energy denominator of the correlation function (see Eq. (27))
is twice the bound state energy $\epsilon_{1/2}$ which
is the energy cost of the polarisation process involving
creation of a `particle-hole' pair.
Finally there is a factor of two corresponding to electron spin. 
The bound state energy  (20) $\epsilon_{1/2}$ is 
 about ${\Delta_0^2 \over 2\epsilon_{F}}$.
The core contribution to the unscreened vortex mass (for parameters
appropriate to the cuprates) is thus estimated to be
$$
m^*_0\simeq {\hbar^2k_{F}^2\over \epsilon_{1/2}} \simeq
100m_e
\eqno(32).
$$
This is larger than  the value calculated above for the vortex mass (Eq. 31)
by a factor of four. This discrepancy arises because the matrix
element in the detailed calculation
 is smaller (by a factor of half approximately)
than the dimensional estimate.

We now consider the effect of Coulomb interactions.
The dipolar charge distribution induced by vortex motion
is screened efficiently by the electrons. This greatly
reduces the vortex kinetic energy.  The 
reduction of the vortex kinetic energy is calculated using
standard self-consistent linear response theory. The
dipolar charge distribution produces an extra electric
potential i.e. changes the electrochemical potential 
of the system.  Any change in the electrochemical
potential causes a change in density which therefore
needs to be calculated self-consistently.  
In the following, we outline the calculation of the reduction
of the vortex mass because of Coulomb screening.
The details are provided in Appendix II.
 Varying the action in Eqs. ( 21) and
(22) with respect to the electric potential $A_0\/$,
we find at the extremum, the Poisson equation
$$\nabla^2 A_0 (\vec{r},z) = -4 \pi e \sum_l
\left<\rho_l (\vec{r}) \right> \delta(z-ld) \eqno{(33)}.$$
We find $<\rho_l (\vec{r}) >\/$, the electron
density, to linear order in $u$ and $A_0$ to be
$$\left<\rho_l(\vec{r}) \right> = e\int d\vec{r'}
\chi^{00}_l (\vec{r}, \vec{r'}) A_0(\vec{r'}, ld) -
 u\int d\vec{r'} \chi^{0x}_l (\vec{r},
\vec{r'}) \eqno{(34)}. $$
In Eq. (34), the density-density and density-current response
functions $\chi^{00}\/$ and $\chi^{0x}\/$ are given by 
$$\chi^{00}_{l} (\vec{r}, \vec{r'} ) = -i \int dt
\theta (t) \left< \left[ \rho_l(\vec{r}, t), \rho_l
(\vec{r'} , 0) \right] \right> \eqno{(35)} $$
and 
$$\chi^{0x}_{l} (\vec{r}, \vec{r'} ) = -i \int dt
\theta (t) \left< \left[ \rho_l(\vec{r}, t), p^{x}_l
(\vec{r'} , 0) \right] \right> \eqno{(36)} $$
The second term on the right hand side of Eq. (34) is
the source term for $<\rho_l>\/$; it has to be determined
by integrating the correlation function $\chi_l^{0x}({\vec r},{\vec {r'}})$
over the co-ordinate ${\vec {r'}}$. This is 
evaluated to be 
$$  \int
\chi^{0 x}_{l} (\vec{r}, \vec{r'} ) d\vec{r'} =
\eta (\vec{r}) \lambda \eqno{(37)}$$
where
$$\eta(\vec{r}) = 2{\sqrt{2\over \epsilon_{1/2}}} f^-_{1/2} (r) f^+_{1/2} (r)
\sin \phi
\eqno(38a)
$$
and
$$\lambda = -\sqrt{2\over\epsilon_{1/2}}\,\,\,\,\hbar 
\int d\vec{r} v^*_{1/2}(\vec{r}) {\partial u^*_{1/2}(\vec{r})
\over \partial x} 
 \eqno{(38b)}
$$
It is thus clear from Eq. (37) substituted into Eq. (34)
that the charge distribution generated by the vortex
motion is proportional to $\sin \phi\/$ i.e. it is dipolar
in nature.  We thus
seek a self-consistent solution to the Poisson equation
of the form 
$$
A_0(\vec{r}, z) = V(r,z) \sin \phi \eqno{(39)}$$
Substituting this form in Eqs. (33) and (34), and using
Eq. (37) we solve for  $A_0({\vec r},z)$ (see Appendix II for details), 
to get
$$
A_0(\vec{q}, ld) = -\frac{4 \pi e u
\lambda \eta (\vec{q})}{ 1 + 2\pi e^2
M(0)} \frac{1}{2q} \left( \frac{ \sinh q d}{\cosh q
d - 1} \right) \eqno{(40)}$$
where $A_0 (\vec{q}, ld)\/$ and $\eta(\vec q)$ are the two dimensional
Fourier transforms of $A_0(\vec{r}, ld)\/$ 
 and $\eta(\vec {r})$ respectively.  In Eq. (40), $M(0)$ is given by 
$$M(0) = \int \frac{d\vec{q}}{(2 \pi)^2} \frac{\mid\eta({\vec q})\mid^2}{2q}
\left( \frac{ \sinh q d}{\cosh q
d - 1} \right)\eqno{(41)}$$
and has the physical significance of an irreducible polarizability .
Knowing the electric potential $A_0
(\vec{q}, ld)\/$ (which is linear in $u$)
we can integrate out the electrons  and the harmonic 
electric potential fluctuations in the  action
(Eq. (22)) to second order in $u$. 
  The result for the extra action per layer is
$$
S_{KE} = \int  dt (u\lambda)^2\left[ 
1 - 
\frac{2\pi e^2
 M(0)}{1+ 2 \pi e^2 M(0)} \right] 
\eqno{(42)}
$$
where the first term in the square bracket is the large 
unscreened contribution
calculated earlier (Eq. (31)), and the cancelling  second term gives the
reduction because of screening.  Combining both these
terms, we find  $m^*\/$ to be
$$
m^* = {m^*_0\over (1+ 2\pi{e^2}M(0))}
 \eqno{(43)}$$

We see from Eq. (43) that there is a `dielectric' screening of the
vortex mass, i.e. the factor in the denominator is a finite large
number. This is due to the discrete level spectrum in the core, in
contrast to the continuum of states in a metal or the near continuum
in conventional superconductors giving `metallic' screening. 
The core dielectric constant ($\epsilon_{core}$) 
can be evaluated using the variational wave functions 
of Eqs. (28) and (30).
The
screening reduces $m^*_0$ by a factor of about 50. 
 The large dielectric constant can be understood as
being due to the high polarizability of the core quasiparticle
system. Approximately, the dielectric constant $(\epsilon_{core})$
 is dimensionally
given by  the ratio of the core Coulomb energy
 and the excitation
energy ($\epsilon_{+-}=2\epsilon_{1/2}$) 
necessary to  create the  particle-hole excitation
which contributes to the polarization process
leading to screening (see Appendix II). Thus,
$$
\epsilon_{\mbox{core}} \simeq
(E_{\mbox{Coulomb}}/E_{\mbox{excitation}}) \simeq \{(e^2/\xi)/
\epsilon_{+-}\} \simeq 75
\eqno(44).
$$
The detailed calculation yields a dielectric constant of 53.  Thus
the screened or effective inertial mass of a vortex is rather small,
being equal to 
$$
m^* = (m_0^*/53) \simeq 0.5 m_e
\eqno(45)$$
This is our main result, for the mass per layer, at $T=0$, when the
field is along the c-axis.

We note that the mass Eq. (45) can be roughly estimated qualitatively
by using the physical estimates Eqs. (32) and (44), i.e.
$$
m^* \simeq \left(\hbar^2 k_F^2/\epsilon_{1/2}\right) \left/
\left\{\left(e^2/\xi\right)/2\epsilon_{1/2}\right\} 
= (2k_F^2a_0\xi) m_e\right .
\eqno(46).
$$
This gives a value $m^* \simeq 1.3 m_e$ 
close to the the result
Eq. (45) of the detailed calculation! It is
clear from the formal expressions, Eq. (29) and Eq. (43) for 
the effective mass, as
well as the approximate form Eq. (46) that $m^*$ is effectively, the
ratio of two polarizabilities, current-current and charge-charge.
Vortex motion (in the rest frame of the vortex) 
 gives rise to both supercurrent fluctuations and
electric potential fluctuations. They are connected because of gauge
invariance.  Since both polarizabilities involve the same energy
denominator, this drops out of $m^*$ (Eq. (46)). It is also clear,
from the occurrence of $k_{F}$, $a_0$ and $\xi$ in this equation, that the
vortex mass is related to the carrier density, core size as well as 
Coulomb interactions.
 The small vortex mass in the cuprates
is a consequence of the small core size and the low carrier density
in comparision with conventional superconductors. 

In the next, concluding section, we discuss the approximations
involved in the result obtained by us, its generalization, and the
question of when vortex mass related effects may be observed.

\noindent {\bf IV.
  Discussion and Conclusion}

\noindent {\bf A. Discussion of the Vortex Mass}
The microscopic calculation above uses an approximate variational
order parameter $\Delta$ and corresponding Bogoliubov-deGennes
amplitudes $(u_{\mu},v_{\mu})$ (Eq. (30)).  An obvious improvement would be to
solve exactly for these quantities given only the parameters
$\Delta_0, \xi$ and the BCS relation $\xi = (\hbar
v_F/\pi\Delta_0)$.  This has been done numerically (22).  There is
of course a fair amount of uncertainty in these parameters for
cuprate superconductors, quite apart from the question of whether an
s-wave, BCS like order parameter with a conventional kinetic energy
functional is at all appropriate for cuprate superconductors.  However,
within the s-wave BCS model, the exact expression Eq. (29) and (43) 
for $m^*$
can be evaluated, once $u_{\mu}$, $v_{\mu}$ and their first spatial
derivative are known for $\mu = 1/2$.   
We have ignored the
contribution due to transitions from the bound states to the
continuum, satisfying the selection rule $\Delta \mu = \pm 1$.  The
reasons are the smallness of the matrix elements, the largeness of
the energy denominator and very good screening.  A rough estimate
shows that these change the mass estimated by about 10-20\%.
We have also ignored  contributions to the mass from polarisation
processes involving the collective excitations of the superconducting state.
In the case of a neutral Fermi superfluid it has been found by Niu et. al.
(40) that the inclusion of these excitations, 
which correspond to long wavelength density
fluctuations, leads to a finite vortex mass in contrast to the logarithmically
divergent result obtained in our approach (see Section I.).
However for the case of a charged superconductor, which is the primary
concern of this paper, the corresponding mode is pushed up to the 
plasma frequency which is much larger than the other energies in the problem
and is therefore unlikely to contribute to the  vortex mass 
in a significant way.

With increasing temperature, the gap $\Delta$({\it T}) and the inverse
coherence length $\xi({\it T})^{-1}$ decrease.  
The structure of the vortex core
also changes.  In principle, one can repeat the $T=0$ calculation with
temperature dependent input parameters. At low temperatures 
$(k_BT\ll \epsilon_{1/2})$  this would roughly have the
effect of {\em increasing} the effective inertial mass of the vortex as
a function of temperature (Eq. 46). 
However, at higher temperatures (when $\epsilon_{1/2}$ is of order or less
than $k_BT$) an 
additional contribution to the vortex mass will accrue from
transitions involving thermally excited quasiparticles in the vortex
core. A qualitatively new effect which arises in this regime 
is that due to quasiparticles scattering off the moving vortex,
there is a damping of vortex motion.
 This dissipative
term is  generally included phenomenologically (it is linear in
vortex velocity u and contributes an imaginary term to the action),
though microscopic theories have been developed (25, 37).  It is not clear
whether the two effects viz. thermal renormalization  of the 
effective mass, and
dissipation, both due to thermal quasiparticles, are completely
independent.   Also, the regime
where the bound core level spacing $\epsilon_{1/2}$ is of order or less
than $k_BT$ is clearly very different from the low temperature regime
$\epsilon_{1/2} >> k_BT$.  We have not considered the former dissipation
dominated `high temperature' limit.

There is the related question of adiabaticity, which is contained in
the assumption that the order parameter of the system with one
moving vortex is the same as that of the static vortex, but with
$\vec{r} \rightarrow (\vec{r}- \vec{u} t)$.  
It is expected that the vortex motion would distort the gap function
from its form in the static case. The distortions induced
have to be determined self-consistently at every
order of $\vec{u}$.
Simanek (26) has
recently considered this question and has pointed out that the vortex
velocity needs to be small enough such that 
$(\hbar u k_{F}) \leq |\epsilon_{1/2}| = (\Delta_0^2/\epsilon_{F})$
or  $u \leq (u_{BCS}) (\Delta_0/\epsilon_F)$ where $u_{BCS}$ is the
BCS critical velocity i.e. $\hbar u_{BCS} k_F \simeq \Delta_0$.  The
violation of this criterion means  that the vortex motion can
significantly distort the gap function as well as the magnetic field and 
supercurrent distribution. The changes induced by the vortex motion
have to be self-consistently determined and it's effects included
in the calculation of the vortex mass. However, at low vortex
velocities ($u\ll u_{BCS}$), 
these effects are small and have therefore been ignored.

We have made a Galilean transformation to the rest frame of the vortex 
to simplify the
calculation of change in action to second order in ${\vec u}$.  
Inclusion of the effects of the periodic lattice and impurity
scattering is also possible within our formalism.
Their interaction potential with the electrons becomes 
explicitly time-dependent and leads to inelastic scattering (38) 
of electrons
whose effects have to be included while evaluating the correlation
functions of Eqs. (24), (35) and (36).

So far our discussion has been restricted to a single
vortex. However, at larger magnetic fields, in the presence
of a vortex lattice,
additional contributions coming from the vortex lattice
will have to be included. The periodicity of the pair potential
broadens the localised quasiparticle levels into energy bands
and new contributions to the vortex mass as well as the forces
experienced by vortices are expected to arise from collective effects
arising from the vortex lattice (15). A totally different
approach becomes necessary in the presence of strong
magnetic fields in the dense vortex limit near $H_{c2}$.
The strong amplitude fluctuations which allow the
dissociation of a Cooper pair make the dynamics of
the order parameter diffusive in this regime.
A calculation of the vortex mass will proceed from the
Abrikosov solution (36) of the GL equations 
for a triangular vortex lattice.
Both the vortex effective mass as  well as the nature
of the forces that drive the vortex dynamics in this
regime are subjects that require further study.

The core contribution to the vortex mass was estimated earlier
microscopically by
by T. Hsu (42).  He obtained an answer which is of the same order of magnitude
as the core contribution calculated by us in the abscence of Coulomb screening
for parameters appropriate to the cuprates.
Hsu used the Bogulibov De-Gennes formalism to obtain the
vortex acceleration 
in response to a transport supercurrent. The vortex mass is then deduced from
this equation by a comparision with a hypothetical 
force equation obtained by setting the unknown vortex mass
times the acceleration of the vortex equal to
 a Magnus force of the size suggested
by Nozieres and Vinen using arguements of fluid hydrodynamics.
While many aspects of the formalism are similar to ours we believe
this method is unreliable for determining the vortex mass.
The Coulomb screening effects which are found to be vital for
giving rise to a small mass have been ignored.
Even for the case of a neutral superfluid, the uncertainty in the size 
as well as sign of the Magnus force
cast doubts about a procedure which relies on a knowledge
of the Magnus force to deduce the core contribution to the vortex mass.

Recently, after completion of this work(15), 
a paper by Simanek (27) appeared  where a  calculation of the
vortex mass using discrete  core states is presented.  This
calculation is based on a TDGL functional, in terms
of a single complex superconducting order parameter,
 which is derived in the
presence of a moving vortex and leads to a 
mass which is dimensionally the same as our estimate
for the unscreened mass.
  However, the all important screening effect which reduces
$m^*$ by a factor of 50-75, has not been considered at all 
 in Simanek's calculation. Further extensions (41) along
 the lines of Ref. (27)  have appeared during the reviewing process.
 However, once again the substantial reduction
in the vortex mass because of Coulomb screening
has been ignored.

The assumption of an s-wave like order parameter is not realistic.  There is
increasing direct experimental evidence for a strongly anisotropic
order parameter, or an order parameter with  vanishing amplitude at
some points in k (or r space) (28,29).  Consider for example a
$d_{x^2-y^2}$ like order parameter.  The pair amplitude is nonlocal
i.e. $<\bar{\psi}_{\uparrow} (\vec{r'}) \bar{\psi}_{\downarrow}
(\vec{r})>$ vanishes for
$\vec{r} = \vec{r'}$, and has a dependence on the direction of
$(\vec{r}-\vec{r'})$ with a square symmetry.  It is thus clear that
the Bogoliubov-deGennes equations are nonlocal mixing different
angular momentum eigenstates. 
Recent work by Volovik (30)  and Ren and co-workers (35) 
(see also Ref. 44 and references therein)
has
suggested that vortices in superconductors with  $d_{x^2-y^2}$
symmetry have a non-zero s-wave component in the core of the
vortex which vanishes at the vortex centre. Thus, the gapless
bound state spectrum, which might have been expected for d-wave 
superconductors with lines of nodes in the gap function (in k-space)
is absent. However, gapless excitations are available in the far
region, where the s-wave component vanishes and are likely to
give rise to strong dissipative effects so that the nature of vortex
dynamics would  be qualitatively
very different.  
This is
an area which needs much further work (see for example ref. 30 and 35). \\

\noindent{\bf B.  Observability of effects due to inertial mass}

It is clear that if there is no
dissipation and no Magnus force, both of which produce a term in the
action linear in velocity (10,11), the small inertial mass of a
vortex would give rise to strong quantum effects.  The vortices are
bosonic particles whose degeneracy temperature is $(eH/m^* c)
(\hbar/k_B)$.  This is of order 20K for an external field of 10T, if the
effective mass $m^*$ is about $0.5 m_{el}$.  If this limit is realized,
then several novel possibilities would arise, especially in strongly
layered cuprate superconductors such as 2212.  In these systems the
vortex liquid phase extends to very low temperatures, especially in
high magnetic fields (31).  This liquid instead of becoming a solid,
could on cooling become a quantum Bose liquid and then a genuine vortex
superfluid (32).  Such a vortex superfluid is a new ground state
with unusual properties, most likely a new kind of insulator.  The
vortex superfluid could persist till $T = 0$, or freeze into a quantum
solid, whose spectrum of collective excitations (phonons) would
depend on the mass $m^*$.  
  The
dynamics of vortices in this regime would be that of interacting 
bosons in a random potential.

It is not clear however that the quantum Bose regime of the many
vortex system is experimentally realizable.  Firstly, at least for
higher temperatures, there is strong dissipation  which dominates
the dynamics in both the quantum and classical regimes.
  Secondly there is a large Magnus   force (9,10,11).  If only
the former were present, the mass could still be relevant for
phenomena like quantum creep.  If only the Magnus force were present, 
as is believed to be the case in the cuprates where the onset of a
dissipationless regime has been reported (5),
the
system of vortices is like that of bosons in a strong `magnetic'
field, the Magnus force being the analogue of the Lorentz force.  The 
Hamiltonian of the system of 2d vortices can be written as
$$
H_{vort} = \sum_i \ (\vec{p}_i - \vec{a}_i)^2/2m^* + (1/2) \sum_{i,j}
V(\vec{r}_i - \vec{r}_{j})
$$
where $(a_x, a_y) = (\pi \hbar n/2) (-x_i, y_i)$ and $V(\vec{r}_i -
\vec{r}_j)$ is the interaction between vortices.  The $\vec{a}$ term
is due to the Magnus force with n being the electron density per
layer and $m^*$ is the vortex effective mass. The `magnetic' field
associated with this Magnus term is rather large, the cyclotron
frequency being about 0.7 eV for $m^* \simeq 0.5 m_e$.  Thus the
Landau level separation is large, and the vortex system is in the
lowest Landau level with a low filling fraction of ($n_v/n)$ where
$n_v$ is the vortex density  and {\it n} the electron density. The magnetic
length of the system, i.e. the cyclotron orbit size is rather small
$\sim$ 7$\AA$ so that the dynamics is that of the guiding centre;
the inertial mass is irrelevant. 
Even in the strong Magnus force limit, a large vortex mass
has been shown (10) to give rise to quantum effects in phenomena
involving vortex tunneling. In particular the semi-classical action
develops a  linear in $T$ dependence, which would reflect in
the observed rate of flux creep at low temperatures.
In the language of our paper, a large vortex mass would
result in a reduction of the cyclotron frequency, mixing
in higher Landau levels and thus enhancing quantum effects.
However, the rather small value of the vortex mass
obtained by us implies that this scenario is actually not realised.
The main uncertainty  in vortex dynamics is the actual
size of the Magnus force.  The contribution of the bound states 
i.e.  of localised quasiparticles to the Magnus force and the effect of
disorder on it are major unsettled issues; there are a number of
suggestions (33,34) that these could reduce, cancel or reverse the Magnus
force! The dynamics of an isolated vortex with inertial mass, in
the presence of a large Magnus force and dissipation  had been investigated
recently (9-12).
 Another possibility is that additional Magnus like
forces
could arise from the pair potential in the dense vortex lattice
limit (15) with an opposite sign.
However, there is a lack of a clear microscopic
theory. There is a growing body of experimental evidence based
on quantum creep (3), Hall measurements (6)  and a.c.
electromagnetic response (14)
that the Magnus force is actually much smaller than current
theoretical estimates (9, 10, 11). 
In that case, there exists the intriguing possibilty of
the formation of a correlated quantum Hall fluid of the bosonic
vortices at low temperatures (39).
With the mounting evidence in the cuprates for a superconducting
order parameter which has $d_{x^2-y^2}$ symmetry
and quasiparticles whose mean free paths could be very long
for $T\ll T_c$,
a realistic picture of this whole field
awaits a microscopic calculation of
the inertial mass, the Magnus force on
a moving vortex  and  dissipation of its momentum 
 for a d-wave superconductor at low temperatures.

\noindent {\bf Acknowledgements}:- We thankfully acknowledge stimulating and
clarifying discussions with B. K. Chakravarty, C. Dasgupta ,
D. Feinberg and H. R. Krishnamurthy.
\newpage
\noindent {\bf Appendix I}

In this appendix we outline the evaluation of the correlation function
(Eq. 24) which determines the unscreened core contribution
to the vortex mass.
Using the Bogulibov transformation (Eq. 25) the field operators
in Eq. (24) can be rewritten in terms of the quasiparticle
operators ($\gamma_{\mu}$) to give
$$m_0^*=\int dt\int d{\vec r}\int d{\vec {r'}
}D({\vec r},{\vec {r'}};t)\eqno {(AI.1)}$$
where the correlation function $D({\vec r},{\vec {r'}};t)$ is given by
$$D({\vec r},{\vec {r'}};t)=\sum_{\mu,\nu,\lambda,\eta}\sum_{\sigma,\sigma'}
f_{\mu}^{\sigma*}({\vec r}){\hbar\over i}{\partial f_{\nu}^{\sigma} ({\vec
r}) \over\partial
x}
f_{\lambda}^{\sigma'*}({\vec {r'}}){\hbar\over i}{\partial 
f_{\eta}^{\sigma'} ({\vec {r'}})\over\partial
x'}\, P_{\mu,\nu}^{\lambda,\eta}(t)
\eqno(AI.2a)$$
where
$$P_{\mu,\nu}^{\lambda,\eta}(t)=i\langle T[\gamma_{\mu}^{\dagger}(t)
\gamma_{\nu}(t)\gamma_{\lambda}^{\dagger}(0)\gamma_{\eta}(0)]\rangle
\eqno(AI.2b)$$.
Here $f_{\mu}^{\uparrow}({\vec r})=u_{\mu}({\vec r})$
and $f_{\mu}^{\downarrow}({\vec r})=v_{\mu}^*({\vec r})$
and the summation with respect to $\mu$, $\nu$, $\lambda$
and $\eta$ runs over both positive and negative values.
The correlation function in Eq. (AI.2) is easily evaluated using 
the diagonalized Hamiltonian (Eq. (26)) to yield 
the expression given in Eq. (27).

We will now derive the selection rule mentioned in Section III.
To find $m_0^*$ from the correlation function of Eq. (27)
we need to integrate with respect to the co-ordinates ${\vec r}$
and ${\vec {r'}}$. The ${\vec r}$ -integration  requires the evaluation
of the integral
$$I_1=\int d{\vec r} u_{\mu}^*({\vec r}){\hbar \over i}{\partial
u_{\mu'}({\vec r})\over
\partial x}\eqno(AI.3)$$
Substituting the explicit forms of $u_{\mu}({\vec r})$ 
and $u_{\mu'}({\vec r})$ (Eq. (28))
we find that
$$I_1={\hbar\over i}\int dr r f_{\mu}^{-}(r){\partial f_{\mu'}^{-}(r)\over 
\partial r}
\int d\phi \cos{\phi} e^{i(\mu-\mu')\phi}$$
$$\,\,\,\,\,\,\,+ 
\hbar(\mu'-1/2)\int dr \, f_{\mu}^{-}(r)f_{\mu '}^{-}(r)
\int d\phi \sin{\phi} e^{i(\mu-\mu')\phi}\eqno (AI.4)$$
The angular integrals in Eq. (AI.4) are zero unless $\mu-\mu'=\pm 1$.
This together with the constraint $\mu>0$,  $\mu'<0$
implies that the only non-zero contribution to $m_{0}^*$
comes from $\mu=-\mu '=1/2$.
Making use of this selection rule, we find, after a little algebra,
that the expression for $m_o^*$ reduces to Eq. (29).
In arriving at this relation, we have
used Eqs. (26b) and (26c).
The only remaining task is to evaluate the matrix element
$g_x$ occuring in Eq. (29).
Substituting the explicit functional forms (Eq. (30)) into
Eq. (29b) we find that
$$g_x={\hbar\over 2\xi}{{\int_0^\infty}dx \,x e^{-x}J_1(k_F\xi x)[
-J_0(k_F\xi x)/2+k_F\xi J_0^{\prime}(k_F\xi x)]\over 
\int_0^\infty dx\, x e^{-x}[J_0^2(k_F\xi x)+J_1^2(k_F\xi x)]}\eqno(AI.5)$$
Evaluating the dimensionless integrals on the R.H.S. of Eq. (AI.5)
we find that for parameters appropriate to the cuprates
($k_F\xi\simeq 4.47$ )
$$\mid g_x\mid\simeq 1.12{\hbar\over\xi}\eqno(AI.6)$$
The largest contribution to $g_{x}$ in Eq. (AI.5) comes from the term
involving $k_F\xi J_0^{\prime}(k_F\xi x)$. However,
unlike conventional superconductors, the relative smallness of the
dimensionless parameter $k_F\xi$ implies that the other  term
 cannot be ignored.
\newpage
\noindent {\bf Appendix II}

In the following, we present details of the calculation
of the large reduction in the core contribution to the vortex
mass due to Coulomb screening. To solve the Poisson equation
(Eqs. (33) and (34)), it is necessary to determine the source
term on the R.H.S. of Eq. (34) by integrating the correlation function
$\chi_l^{0x}({\vec r},{\vec{r'}})$ over the co-ordinate ${\vec {r'}}$.
Making use of the Bogulibov transformation ( Eq. (25))
and the diagonalised Hamiltonian (Eq. (26)) we find
(after some algebra) that
$$\chi_l^{0x}({\vec r},{\vec{r'}})
= 2 \sum^{unocc.}_{\mu>0}
\sum^{occ.}_{\nu<0} {u^*_\mu (\vec{r}) 
u_{\nu} (\vec{r})\over (\epsilon_\mu -
\epsilon_{\nu})}[v_{\mu }(\vec{r'})
{\hbar\over i} {\partial v_{\nu}^*(\vec {r'})\over \partial x'
} -u^*_{\nu }(\vec{r'})
{\hbar\over i} {\partial u_{\mu}(\vec{r'})\over \partial x'
} ]+h.c. \eqno(AII.1)$$

The operator ${\hbar\over i}{\partial \over \partial x}$
behaves like $\cos\phi$ ($\sin\phi$) and as before,
the integration over the co-ordinate ${\vec{r'}}$
(see Eq. (AI.4) and the discussion following it)
gives the selection rule $\mu-\nu=\pm 1$
which together with the constraint $\mu > 0$
and $\nu < 0$ implies $\mu=-\nu=1/2$.
We therefore find
$$\int \chi^{0 x}_{l} (\vec{r}, \vec{r'} ) d\vec{r'} =
{-2u_{1/2}^*({\vec r})v_{1/2}^*({\vec r})\over 
\epsilon_{1/2}}\int d{\vec {r'}} v_{1/2}({\vec{r'}})
{\hbar\over i}{\partial u_{1/2}(\vec {r'})
\over \partial x'}
+ h.c. \eqno(AII.2)$$

In writing Eq. (AII.2) we have used Eqs. (26b)
and (26c). Now substituting the explicit variational 
forms for
$u_{1/2}({\vec r})$ and $v_{1/2}({\vec r})$ (Eqs. (28)
and (30)) we arrive at Eqs. (37) and (38).
To proceed further, we have to solve the Poisson equation
and get a self-consistent solution for the scalar potential
$A_0({\vec r},z)$ of the form assumed in Eq. (39).
We will now consider the other term on the R.H.S.
of Eq. (34). This term represents the screening charge
induced by the Coulomb potential consequent
to the electron density change induced by the vortex motion;
it has to be determined by integrating the product
of $\chi^{00}_{l} (\vec{r}, \vec{r'} )$and $A_0({\vec{r'}},ld)$
with respect to the co-ordinate ${\vec{r'}}$.
Since the latter has an angular dependence of the form
$\sin \phi$ we once again find that the only process
which contributes to the polarization  involves a
transition from the highest occupied state ($\nu=-1/2$)
to the lowest unoccupied state ($\mu=1/2$).
We thus find,
$$\int d{\vec{r'}}\chi^{00}_l({\vec r},{\vec{r'}})A_0({\vec{r'}},ld)
=-{\eta({\vec r})\over 2}\int d{\vec {r'}} A_0({\vec{r'}},ld) \eta(\vec{r'})
\eqno(AII.3)$$
Combining Eqs. (37) and (AII.3) with the Poisson equation
(Eqs. (33) and (34)) we get
$${\nabla ^2 A_0({\vec r},z)\over 4\pi} =
e\eta({\vec r})\sum_{l} \delta(z-ld)[u\lambda + {e\over 2}\int d{\vec {r'}}
\eta({\vec{r'}})A_0({\vec{r'}},ld)]\eqno(AII.4)$$
Transforming to Fourier space, this can be rewritten
in terms of the corresponding Fourier components as
$${[-q^2-k^2]A_0({\vec q},k)\over 4\pi}= 
eu\lambda\eta(\vec{q})\sum_l \exp{[-ikld]} + {e^2\eta({\vec q})\over 2d}
\sum_m\int {d{\vec{q'}}\over (2\pi)^2}\eta(-{\vec{q'}})A_0({\vec{q'}},k-{2
\pi m\over d})\eqno(AII.5)$$
This is an integral equation for the scalar potential
$A_0$. To solve for $A_0$, we find it convenient
to introduce the quantity
$$X(k)={1\over d}\sum_m\int {d{\vec q}\over (2\pi)^2}\eta(-{\vec q}) 
A_0({\vec q},k-{2\pi m\over d})\eqno(AII.6)$$
Substituting Eq. (AII.6) in (AII.5)  and making use of the property
$X(k)=X(k-{2\pi m\over d})$ for any integer $m$,
we solve for $X(k)$ to get
$$X(k)={-4\pi eu\lambda M(k)\sum_l e^{-ikld}\over 
1+2\pi e^2 M(k)}\eqno(AII.7)$$
where 
$$M(k)=\int \frac{d\vec{q}}{(2 \pi)^2} \frac{\mid\eta({\vec q})\mid^2}{2q}
\left( \frac{ \sinh q d}{\cosh q
d - \cos k d} \right)\eqno(AII.8)$$

Substituting Eqs. (AII.7) and (AII.8) in Eq. (AII.5)
we can now solve for $A_0({\vec q},k)$ to get
$$A_0({\vec q},k)={-4\pi eu\lambda\over {q^2+k^2}}
{\eta({\vec q})\over {1+2\pi e^2 M(k)}}\sum_l \exp{(-ikld)}\eqno(AII.9)$$
Fourier transforming the above equation
with respect to the wave vector $k$ we 
get   $A_0({\vec q},ld)$ (Eq. (40)).

We now consider the action Eq. (22). We first consider
the electric field energy. Integrating by parts,
and making use of the Poisson equation 
(Eq. (33)) we get 
$$\int dz\int d{\vec r} {(\nabla A_0({\vec r},z))^2\over 8\pi}
={-e\over 2}\sum_l\int d{\vec r}\int d{\vec {r'}}A_0({\vec r},ld)
[u\chi^{0x}_l({\vec r},{\vec {r'}})-e\chi^{00}_l({\vec r},{\vec {r'}})
A_0({\vec {r'}},ld)]\eqno(AII.10)$$
On integrating out the electrons to second order in 
the vortex velocity $u$ and the Coulomb potential $A_0$
we get the effective action
$$S^{\prime}=\sum_l\int dt\int d{\vec r}d{\vec {r'}}[{-u^2\over 2}
\chi^{xx}_l({\vec r},{\vec {r'}})\,-\,{e^2\over 2}A_0({\vec r},ld)
\chi^{00}_l({\vec r},{\vec {r'}})A_0({\vec {r'}},ld) \,
+\,ueA_0({\vec r},ld)\chi^{0x}_l({\vec r},{\vec {r'}})]
\eqno(AII.11)$$

Combining Eqs. (AII.10) and (AII.11) we find that the effective
action for the system, to second order in $u$, is given by
$$S_{KE}=\sum_l\int dt\int d{\vec r}d{\vec {r'}}[{-u^2\over 2}
\chi^{xx}_l({\vec r},{\vec {r'}})
+\,{ue\over 2}A_0({\vec r},ld)\chi^{0x}_l({\vec r},{\vec {r'}})]
\eqno(AII.12)$$
The first term in $S_{KE}$ is the unscreened core contribution
to the vortex mass evaluated earlier (Eq. (24)) while the other 
term represents the reduction because of Coulomb screening.
We now substitute the explicit forms for $\chi_l^{xx}$
and $\chi_l^{0x}$.
Doing a calculation very similar to the one leading
to Eqs. (37), (38) and (AII.2) we find
$$\int d{\vec r}d{\vec {r'}}
\chi^{xx}_l({\vec r},{\vec {r'}})=-2\lambda^2\eqno(AII.13)$$
Substituting Eqs. (37) and (AII.13)  into  Eq. (AII.12)
we get
$$S_{KE}=\sum_l\int dt[u^2\lambda^2+{eu\lambda \over 2}\sum_l\int 
d{\vec r}\eta({\vec r})A_0({\vec r},ld)]\eqno(AII.14)$$
Fourier transforming the second term in Eq. (AII.14)
and substituting the expression for $A_0$ (Eq. (40))
we finally arrive at Eq. (42).

The only remaining task is to evaluate the polarizability
$M(0)$. Using Eq. (38a), we find that
$$\eta({\vec q})=2\sqrt{2\over\epsilon_{1/2}}\int d{\vec r}
f^-_{1/2}(r)f^+_{1/2}(r)\sin\phi e^{-i{\vec q}\cdot{\vec r}}\eqno(AII.15)$$
On substituting the variational forms (Eq. 30) for $f^-_{1/2}(r)$
and $f^+_{1/2}(r)$ and integrating over the angular co-ordinate
$\phi$ this reduces to
$$\eta({\vec q})=2\sqrt{2\over\epsilon_{1/2}}{2\pi i\over A_{1/2}^2}
{q_y\over q}\int_0^{\infty} dr r J_0(k_Fr)J_1(k_Fr)J_1(qr)
e^{-r/\xi}\eqno(AII.16)$$
Combining Eq. (AII.16) with the expression for $M(0)$
(Eq. (41)) we finally get
$$M(0)={4\pi\over A_{1/2}^4\epsilon_{1/2}}\int dq {\sinh (qd)\over
\cosh (qd)-1}I^2(q) \eqno(AII.17)$$
where
$$I(q)=\int_0^{\infty} dr r J_0(k_Fr)J_1(k_Fr)J_1(qr)e^{-r/\xi}
\eqno(AII.18)$$

Using Eqs. (30d), (43), (AII.17) and (AII.18) we are now
in a position to calculate the `core dielectric constant'
$\epsilon_{core}$. We find that
$\epsilon_{core}=1+2\pi e^2M(0)$ is given by
$$\epsilon_{core}=1+{e^2/\xi\over \epsilon_{+-}}{L_1\over L_2^2}
\eqno(AII.19)$$
where 
$$L_1=4\int_0^{\infty}dx {\sinh(xd/\xi)\over \cosh(xd/\xi)-1}f^2(x)
\eqno(AII.20)$$
$$L_2=\int_0^{\infty}dx\,x e^{-x}(J_0^2(k_F\xi x)+J_1^2(k_F\xi x))
\eqno(AII.21)$$
and
$$f(x)=\int_0^\infty dy y J_0(k_F\xi y)J_1(k_F\xi y)J_1(xy)e^{-y}
\eqno(AII.22)$$
On evaluating these expressions (Eqs. (AII.19), (AII.20) and 
(AII.21) numerically we find $\epsilon_{core}\simeq 53$.

\newpage
\noindent{\bf Figure Captions}

{\bf Fig. 1} Feynman diagrams that contribute to the phase-only
action functional of Eq. 1)  at $T=0$ in the clean limit 
on integrating out the electrons.
The thick lines correspond to the fermions, the wavy lines indicate the
density fluctuations
$\rho=(\hbar{\dot\theta}-2eA_0)$ and the dashed line stands for the supercurrent
fluctuations ${\vec j}=(\hbar\nabla\theta-2e{\vec A}/c)$.

{\bf Fig. 2} Additional diagrams that contribute to the Galilean invariant
action functional of Eq. 7) on integrating out the electrons.
The thick lines correspond to the fermions, the wavy lines indicate the
density fluctuations
$\rho=(\hbar{\dot\theta}-2eA_0)$ and the dashed line stands for the supercurrent
fluctuations ${\vec j}=(\hbar\nabla\theta-2e{\vec A}/c)$.

{\bf Fig. 3} Spectrum of bound states within the core (within the range $-\Delta_0$
to $\Delta_0$) and of continuum states outside it, with the angular
momentum quantum numbers. The levels are appropriate to the
parameters mentioned in the text for a cuprate superconductor at 
$T=0$. The transition allowed by selection rules is shown by an
arrow. 
\newpage
\noindent {\bf References}

\begin{enumerate}

\item G. Blatter, M.V. Feigel'man, V.B. Geshkenbein, A.I. Larkin and
V.M. Vinokur, {\em Rev. Mod. Phys.} {\bf 66}, 1125 (1994).

\item A.C. Mota,  Pollini, P. Visani, K.A. Muller and J.G.
Bednorz,  {\em Physica Scripta} {\bf 37}, B23 (1988).
L. Fruchter et. al. {\em Phys. Rev.} {\bf B43}, 8709 (1991).
D. Prost et. al. , {\em Phys. Rev.} {\bf B47}, 3457 (1993).
\item G.T. Seidler, T.F. Rosenbaum, K.M. Beauchamp, H.M. Jaeger,
G.W. Crabtree and V.M. Vinokur, {\em Phys. Rev. Lett.} {\bf 74}, 1442
(1995).

\item M. Galffy and E. Zirngiebl, {\em Solid State Comm.} {\bf 68},
929 (1988). \\ 
S.J. Hagen, C.J. Lobb, R.L. Greene, M.J. Forester and J.H. Kang,
{\em Phys. Rev.} {\bf B41}, 11630 (1990). \\
T.R.  Chin, T.W. Jing, N.P. Ong, and Z. Z. Wang, {\em Phys. Rev. Lett.}
{\bf 66}, 3075 (1991).

\item J.M. Harris, Y.F. Yan, O.K.C. Tsui, Y. Matsuda and N.P. Ong,
{\em Phys. Rev. Lett.} {\bf 73}, 1711 (1994).

\item S. Bhattacharya, M.J. Higgins and T.V. Ramakrishnan, {\em Phys. Rev.
Lett.} {\bf 73}, 1699 (1994).

\item S. Spielman et. al., {\em Phys. Rev. Lett.} {\bf 73}, 1537 (1994).

\item R. Theron et al., {\em Phys. Rev. Lett.} {\bf 71}, 1246 (1993).

\item M.V. Feigel'man, V.B. Geshkenbein, A.I. Larkin and S. Levit,
{\em JETP Lett.} {\bf 57}, 711 (1993).

\item P. Ao and D.J. Thouless, {\em Phys. Rev. Lett.}{\bf 70}, 2158 (1993);
{\it ibid} {\bf 72}, 132 (1994).

\item M.J. Stephen, {\em Phys. Rev. Lett.} {\bf 72}, 1534 (1994).

\item C. Morais-Smith, B. Ivlev and G. Blatter {\em Phys. Rev.} B {\bf
49}, 4033, (1994).

\item G. Blatter, V.B. Geshkenbein and V.M. Vinokur, {\em Phys. Rev.
Lett.} {\bf 66}, 3297 (1991).

\item B. Parks et. al. {\em Phys. Rev.
Lett.} {\bf 74}, 3625  (1995).

\item D.M. Gaitonde and T.V. Ramakrishnan, {\em Physica C} {\bf
235-40}, 245 (1994).

\item H. Suhl, {\em Phys. Rev. Lett.} {\bf 14}, 226 (1965).

\item J-M. Duan and A.J. Leggett, {\em Phys. Rev. Lett.} {\bf 68},
1216 (1992).

\item J-M. Duan {\em Phys. Rev.} B {\bf 48}, 333 (1993).

\item T. V. Ramakrishnan, {\em Physica Scripta} T {\bf 27}, 24 (1989).

\item C. Caroli, P.G. de Gennes and J. Matricon, {\em Phys. 
Lett.} {\bf 9}, 307 (1964).

\item J. Bardeen, R. Kummel, A.E. Jacob and L. Tewordt, {\em Phys.
Rev.} {\bf 187}, 556 (1969).

\item F. Gygi and M. Schluter, {\em Phys. Rev.} B. {\bf 43}, 7609 (1991).

\item Y-D Zhu, F-C Zhang and H.D. Drew, {\em Phys. Rev.} B {\bf
47}, 587 (1993).

\item  I. J. R. Aitchison et. al., {\em Phys. Rev.} B {\bf 51}, 6531 (1995).

\item N.B. Kopnin and V.F. Kravtsov, {\em Sov. Phys. JETP} {\bf 44},
861 (1976).

\item E. Simanek, {\em Phys. Rev.} B {\bf 46}, 14054 (1992).

\item E. Simanek {\em Phys. Lett.} A {\bf 194}, 323 (1994).

\item D.J. Scalapino, {\em Physics Reports} {\bf 250}, 329 (1995).

\item Z. X. Shen, W. E. Spicer, D. M. King, D. S. Dessau, B. O. Wells
{\em Science
} {\bf 267}, 343 (1995).

\item G.E. Volovik, {\em J.E.T.P. Lett.} {\bf 58}, 469  (1993).

\item  D. J. Bishop, P. L. Gammel, D. A. Huse and C. A. Murray,
{\em Science} {\bf 255}, 165, (1992).
\item D. H. Lee and M. P. A. Fisher, {\em Int. J. Mod. Phys. }
{\bf B5}, 2675 (1991).
\item G. E. Volovik  {\em J.E.T.P.} {\bf 77}, 435 (1993).
\item M. V. Feigel'man, V. B. Geshkenbein, A. I. Larkin and
V. M. Vinokur (ETH preprint 1995).
\item Y. Ren, J-H. Xu and C. S. Ting {\em Phys. Rev. Lett.}
{\bf 74}, 3680 (1995).
\item A. Abrikosov, Fundamentals of the Theory of Normal Metals,
North-Holland (1988)
\item D. M. Gaitonde and T. V. Ramakrishnan (to be published).
\item F. Guinea and Yu Pogorelov, {\em Phys. Rev. Lett.}
{\bf 74}, 462 (1995).
\item M. Y. Choi, {\em Phys. Rev. }B
{\bf 50}, 10,088 (1995); Ady Stern {\em Phys. Rev.} B {\bf 50}, 10,092 (1995).
\item Q. Niu et. al. {\em Phys. Rev. Lett.} {\bf 72}, 1706 (1994).
\item E. Simanek, {\em Jour. of Low Temp. Phys.} {\bf 100}, 1 (1995); 
A. V. Otterlo et. al. {\em Phys. Rev. Lett.} {\bf 75}, 3736 (1996).
\item T. C. Hsu {\em Physica} {\bf C213}, 305 (1993).
\item Ch. Renner, I. Maggio-Aprile, A. Erb, E. Walker
 and O. Fischer (to be published).
\item J. Hai, Y. Ren and C. S. Ting, {\em Int. Jour. of Mod. Phys.}
{\bf B10}, 2699 (1996).
\end{enumerate}

\end{document}